\documentclass[twoside]{article}
\usepackage{amsmath}
\begin{document}
\begin{center}
{\bf Schwarzschild solution in R-spacetime}
\end{center}
\begin{center}
{T.~Angsachon, S.~N.~Manida}
\end{center}
\begin{center}
{\it Department of High Energy Physics, Faculty of Physics,\\ Saint-Petersburg State University, Russia.}
\end{center}
Here we construct new solutions from the Einstein Field equations --
some analog of the Schwarzschild metric in anti--de~Sitter--Beltrami spacetime in the $c\to \infty$ limit ($R$-spacetime). In this case we derive an adiabatic invariant for finite movement of the massive point particle and separate  variables Hamilton--Jacobi equation.
Quasi orbital motion is analyzed and its radius time dependence  is obtained.

\section{Schwarzschild metric in Anti de-Sitter-Beltrami spacetime}

One of the important solution of Einstein's Field Equations - this is the Schwarzschild solution, which describes a gravitational field
outer the massive spherical symmetric body. If the spacetime is a large distance form a source of this field with negative constant curvature R
(Anti de-Sitter spacetime), then corresponding metric in a spherical coordinate system  ${x_0,r,\theta,\varphi}$ is written out as \cite{S}:
\begin{equation}
ds^2 = \left(1+\frac{r^2}{R^2}-\frac{2M}{r}\right)dx_0^2-\frac{dr^2}
{\left(1+\displaystyle\frac{r^2}{R^2}-\displaystyle\frac{2M}{r}\right)}-r^2d\Omega^2, \label{01}
\end{equation}
where $d\Omega^2=d\theta^2+\sin^2\theta d\varphi^2$.
This is the Schwarzschild-Anti de-Sitter metric $(SAdS)$.
Now we change the metric (\ref{01}) to the Beltrami coordinates. To do that we introduce coordinate transformations \cite{M2}:
\begin{equation}
x_0\Rightarrow R\arcsin \dfrac{x_0}{R\sqrt{1+\dfrac{x_0^2}{R^2}}},\,\,\,\,\,
r\Rightarrow \dfrac{r}{\sqrt{1+\dfrac{x_0^2}{R^2}-\dfrac{r^2}{R^2}}}.
\label{02}
\end{equation}
To put these transformations (\ref{02}) in metric (\ref{01}), we get that
\begin{equation}
ds^2=\dfrac{dx^2}{h^2}-\dfrac{(xdx)^2}{R^2h^4}-\dfrac{2Mh}{rh_0^4}dx_0^2
-\dfrac{2M\left(dr-\dfrac{rx_0dx_0}{R^2h_0^2}\right)^2}{rh\left(1-\dfrac{2Mh^3}{rh_0^2}\right)}.
\label{03}
\end{equation}
For abbreviation of formulas we define:  $dx^2\equiv dx_0^2-d\vec{r}^2,\,\,xdx\equiv x_0dx_0-\vec{r}d\vec{r},\,\,h^2\equiv 1+x^2/R^2,\,\, h_0^2\equiv 1+x_0^2/R^2.$

A formula (\ref{03}) is the Schwarzschild-Anti de-Sitter metric in the Beltrami coordinates (SAdSB).

Now this metric is changed into the limit $c\rightarrow\infty$ and we define  $Mc\equiv g$, so the metric (\ref{03}) will be
\begin{equation}
ds^2=\dfrac{R^4}{c^2t^4}\left(1-\dfrac{2gt}{rR}\right)dt^2-\dfrac{R^2(tdr-rdt)^2}
{c^2t^4\left(1-\dfrac{2gt}{rR}\right)}-\dfrac{R^2r^2}{c^2t^2}d\Omega^2.
\label{04}
\end{equation}
This is an approach expression for the metric $SAdSB$ which is called that the Schwarzschild solution in $R$-spacetime.

We can write out an action for classical test particle with mass $m:$
\begin{equation}
S = -mc\int{}ds = -mR^2\int{}\frac{dt}{t^2}\sqrt{1-\frac{2gt}{rR}
-\frac{(t\dot{r}-r)^2}{R^2\left(1-\dfrac{2gt}{rR}\right)}-\frac{r^2t^2\dot{\varphi}^2}{R^2}},
\label{05}
\end{equation}
where $\dot{r}={dr}/{dt}$,$\dot{\varphi}={d\Omega}/{dt}.$

Lagrangian of this system has a form
\begin{equation}
L = -m\frac{R^2}{t^2}\sqrt{1-\frac{2gt}{rR}-\frac{(t\dot{r}-r)^2}
{R^2\left(1-\dfrac{2gt}{rR}\right)}-\frac{r^2t^2\dot{\varphi}^2}{R^2}}.
\label{06}
\end{equation}
The research about the properties of this Lagrangian will be provided in next part.

\section{The limit of classical mechanics in R-spacetime}

To get the classical mechanics limit we consider a spacetime region $gt/R\ll r\ll R$, $\dot{r}\ll R/t$ and expand
Lagrangian (\ref{06}) in the order of the small parameters $r/R$, $\dot{r}t/R$, $gt/(rR):$\\
\begin{equation*}
L \simeq -m\dfrac{R^2}{t^2}\left(1-\dfrac{1}{2}\left(\dfrac{2gt}{rR}+\dfrac{(t\dot{r}-r)^2}{R^2\left(1-\dfrac{2gt}{rR}\right)}
+\dfrac{r^2t^2\dot{\varphi}^2}{R^2}\right)\right)
\end{equation*}
\begin{equation}
= -m\frac{R^2}{t^2}+\dfrac{mgR}{rt}+\dfrac{m(t\dot{r}-r)^2}{2t^2}+\dfrac{mr^2\dot{\varphi}^2}{2}. \label{07}
\end{equation}
Because of the equations of motion do not change under the addition of some coordinate and time function into the Lagrangian function,
and (\ref{07}) will be written as
\begin{equation}
L= \frac{m\dot{r}^2}{2}+\frac{mr^2\dot{\varphi}^2}{2}+\frac{mgR}{rt} +\frac{d}{dt}\left(\frac{R^2}{t}-\frac{r^2}{2t}\right),
   \label{08}
\end{equation}
and the total derivative have been neglected.

Now we transform to the Hamiltonian formalism and introduce together the radial momentum
\begin{equation}
p_r = \frac{\partial{}L}{\partial\dot{r}} = m\dot{r}
\label{09}
\end{equation}
and the angular momentum
\begin{equation}
p_\varphi  = \frac{\partial{}L}{\partial\dot{\varphi}} = mr^2\dot{\varphi}\equiv  M_\varphi,
\label{10}
\end{equation}
which is the integral of motion.

The Hamiltonian function will be constructed with the help of the Legendre transform
\begin{equation}
H = p_r\dot{r}+p_\varphi\dot{\varphi}-L = \frac{p_r^2}{2m}+\frac{p_\varphi^2}{2mr^2}-\frac{mgR}{rt}. \label{11}
\end{equation}
A last term in (\ref{11}) can be considered as the gravitational potential
$$V= -\dfrac{mG(t)}{r}$$
with slowly changed gravitational constant
$$G(t)=\dfrac{gR}{t}.$$
Obviously that the Hamiltonian (\ref{11}) contains a parameter which exactly, but slowly depends on time,
therefore we use the adiabatically invariant method to solve this problem\cite{L}.

The stationary Hamiltonian function of our system is represented as
\begin{equation}
H_0 = \frac{p_r^2}{2m}+\frac{p_\varphi^2}{2mr^2}-\frac{mG_0}{r} = E. \label{12}
\end{equation}
The Hamilton-Jacobi equation is derived in a simple form
\begin{equation}
\frac{1}{2m}\left(\frac{\partial{}S}{\partial{}r}\right)^2+
\frac{1}{2mr^2}\left(\frac{\partial{}S}{\partial\varphi}\right)^2-\frac{mG_0}{r} = E_0,
\label{13}
\end{equation}
where $S$ is a completed action for considered system. We will find it in the formula
\begin{equation}
S = S_r(r)+M_\varphi\varphi-E_0t.\label{14}
\end{equation}
We put the formula (\ref{14}) in the equation (\ref{13}) and get that
\begin{equation}
\frac{\partial S_r(r)}{\partial r} = \sqrt{2mE_0+\frac{2m^2G_0}{r}-\frac{M_\varphi^2}{r^2}} = p_r. \label{15}
\end{equation}
Therefore the adiabatically invariants will be in these formulations
\begin{equation}
I_r = \frac{1}{2\pi}\oint{}p_rdr =-M_\varphi+G_0\sqrt{\frac{m^3}{2|E_0|}},
\label{16}
\end{equation}
\begin{equation}
I_\varphi=\frac{1}{2\pi}\oint{}p_\varphi\varphi = M_\varphi
\label{17}
\end{equation}
From the sum of (\ref{16}) and (\ref{17}) following that the formulation for energy's system and dependent on time
\begin{equation}
E = -\frac{m^3 g^2R^2}{2(I_r+I_\varphi)^2 t^2}.
\label{18}
\end{equation}
\section{The Schwarzschild metric in R-spacetime and the motion on the quasi-circular orbits}

In this part we consider the motion of a test point particle with mass m in the Schwarzschild metric in R-spacetime (\ref{04}).
To do that we use the Hamilton-Jacobi method. In General Relativity Theory the Hamilton-Jacobi equation can be represented as \cite{L2}
\begin{equation*}
g^{\mu\nu}\frac{\partial{}S}{\partial{}x^\mu}\frac{\partial{}S}{\partial{}x^\nu} = m^2c^2
\end{equation*}
Furthermore the Hamilton-Jacobi equation for a metric in R-spacetime is expressed in a formula
\begin{multline}
\left[\frac{t^2}{R^2}\frac{1}{\sqrt{1-\dfrac{2gt}{rR}}}\frac{\partial{}S}{\partial{}t}+\frac{tr}{R^2}\frac{1}
{1-\dfrac{2gt}{rR}}\frac{\partial{}S}{\partial{}r}\right]^2
-\frac{t^2}{R^2}\left(1-\frac{2gt}{rR}\right)\left(\frac{\partial S}{\partial r}\right)^2-\\
-\frac{t^2}{R^2r^2}\left[\left(\frac{\partial{}S}{\partial\theta}\right)^2+\frac{1}{\sin^2\theta}\left(\frac{\partial{}S}
{\partial\varphi}\right)^2\right]= m^2. \label{21}
\end{multline}
To consider a motion on the surface $\theta = \dfrac{\pi}{2}$ the Hamilton-Jacobi equation (\ref{21}) is simplified
\begin{multline}
\left[\frac{t^2}{R^2}\frac{1}{\sqrt{1-\dfrac{2gt}{rR}}}\frac{\partial{}S}{\partial{}t}+\frac{tr}{R^2}\frac{1}
{1-\dfrac{2gt}{rR}}\frac{\partial{}S}{\partial{}r}\right]^2
-\frac{t^2}{R^2}\left(1-\frac{2gt}{rR}\right)\left(\frac{\partial{}S}{\partial{}r}\right)^2-\\
-\frac{t^2}{R^2r^2\sin^2\theta}\left(\frac{\partial{}S}
{\partial\varphi}\right)^2= m^2.\label{22}
\end{multline}
To solve this equation we use the method of separation o variables, which contains that a completed action
can be as the sum of each variable functions
\begin{equation}
S = S_1(t)+S_2\left(\frac{r}{t}\right)+M_\varphi\varphi. \label{23}
\end{equation}
When we have put the separation of variables (\ref{23}) in the equation (\ref{22}) and we can find that
\begin{equation*}
S_1 = \frac{-A}{t}
\label{24}
\end{equation*}
and
\begin{equation*}
S_2 = \int{}dx\sqrt{\frac{A^2}{R^2}\frac{1}{\left(1-\dfrac{x_g}{x}\right)^2}
-\frac{m^2R^2+\dfrac{M_\varphi^2}{x^2}}{1-\dfrac{x_g}{x}}},
\label{25}
\end{equation*}
where $x\equiv{r}/{t}$, $x_g \equiv{2g}/{R},$ $A$ is a constant of variable separation.
Finally the action is written out as
\begin{equation}
S = \frac{-A}{t}+\int{}dx\sqrt{\frac{A^2}{R^2}\frac{1}{\left(1-\dfrac{x_g}{x}\right)^2}
-\frac{m^2R^2+\dfrac{M_\varphi^2}{x^2}}{1-\dfrac{x_g}{x}}}+M_\varphi\varphi. \label{26}
\end{equation}
Two equations of motion are derived from the derivative (\ref{26}) of two constants $A$ and $M_\varphi$:
\begin{equation}
\dfrac{\partial S}{\partial A} = {\text{const}} \label{x}
\end{equation}
and
\begin{equation}
\dfrac{\partial S}{\partial M_\varphi} = {\text{const}}.\label{y}
\end{equation}
From an equation (\ref{x}) we can find the dependence  $r(t)$
\begin{equation}
\frac{R}{t} = \frac{A}{m}\int\frac{dx}{\left(1-\dfrac{x_g}{x}\right)\sqrt{\dfrac{A^2}{m^2}-
\left(1+\dfrac{M_\varphi^2}{m^2R^2x^2}\right)\left(1-\dfrac{x_g}{x}\right)}},\label{27}
\end{equation}
and an equation of trajectory can be found from the equation (\ref{y})
\begin{equation*}
\varphi = \int{}dx\frac{M_\varphi}{x^2\sqrt{\dfrac{A^2}{R^2}-
\left(\dfrac{M_\varphi^2}{x^2}+m^2R^2\right)
\left(1-\dfrac{x_g}{x}\right)}}.\label{28}
\end{equation*}
We consider now a motion of the test particle on almost the circular orbit. To derive that we rewrite equation (\ref{27})
in differential form
\begin{equation*}
\frac{1}{1-\displaystyle\frac{x_g}{x}}\frac{dx}{dt} = \frac{-R}{At^2}\sqrt{A^2-U^2(x)},\label{29}
\end{equation*}
where
\begin{equation*}
U(x) = mR^2\sqrt{\left(1-\dfrac{x_g}{x}\right)\left(1+\dfrac{M_\varphi^2}{m^2R^2x^2}\right)}
\label{29a}
\end{equation*}
which play a role of the effective potential.

A motion on a circular orbit occurs under following conditions
\begin{equation}
A = U, \,\,\,\,\frac{dU}{dx} = 0. \label{30}
\end{equation}
A solution of the equation (\ref{30}) is represented in this formula
\begin{equation}
r_{\pm} = \frac{1}{2}\frac{M_\varphi^2}{m^2g^2}\frac{gt}{R}\left(1\pm\sqrt{1-\frac{3m^2R^2x_g^2}{M_\varphi^2}}\right). \label{31}
\end{equation}
To easy see that $r_{+}$ is responded to a stable "circular" orbit and $r_{-}$ to a nonstable orbit.

Multiplier  $gt/R$  in the formula is a mean that the Schwarzschild radius. Under the limit $r \gg gt/R$
under-quadratic term is nearly to one and radius become
\begin{equation*}
r_{+}\simeq \frac{M_\varphi^2}{m^2g^2}\left(\frac{gt}{R}\right). \label{33}
\end{equation*}
If we put $t = T+\tau$, where $T$ is the Cosmic time, then the circular orbit radius will be depended on $\tau:$
\begin{equation}
r_+(\tau) = r_+(0)\left(1+\frac{\tau}{T}\right). \label{34}
\end{equation}
For example we consider the change of Moon radius in one year. The Moon orbital radius $r_m\simeq 4\cdot 10^8$~m,
The cosmological time $T\simeq 5\cdot 10^{17}$~c, year period $\tau\simeq 2,5\cdot10^8$~c.
From the equation (\ref{34}) we get the year rising orbital radius of Moon $\Delta r\simeq 2$~mm/year.
This change almost is corresponding with the observation of the rising orbital radius of the Moon (34~mm/year)
which is caused by Tidal force and breaking of Earth's rotation.

\end{document}